\documentclass[twocolumn,amssymb,amssymb,aps,pra,showpacs]{revtex4}
\usepackage{epsfig}
\usepackage{amsfonts}
\usepackage{amsmath}
\usepackage{amsthm}

\begin{document}

\title{Static charged fluid around a massive magnetic dipole}

\author{Jos\'e D. Polanco} \email[e-mail: ]{pepediaz@ifi.unicamp.br} 
\altaffiliation{Present address: Departamento de Ciencias F\'isicas y Matem\'aticas, 
Universidad Arturo Prat, Casilla 121, Iquique, Chile}

\affiliation{Instituto de F\'{\i}sica `Gleb Wataghin', Universidade
Estadual de Campinas - UNICAMP, 13083-970, Campinas, S\~ao Paulo, Brasil}

\author{Patricio S. Letelier}\email[e-mail: ]{letelier@ime.unicamp.br}

\affiliation{Departamento de Matem\'atica Aplicada, Instituto de
Matem\'atica, Estat\'{\i}stica e Computa\c{c}\~ao Cient\'{\i}fica,
Universidade Estadual de Campinas - UNICAMP, 13081-970, Campinas, S\~ao Paulo, 
Brasil}

\author{Maximiliano Ujevic}\email[e-mail: ]{mujevic@ufabc.edu.br}

\affiliation{Centro de Ci\^encias Naturais e Humanas, Universidade 
Federal do ABC, 09210-170, Santo Andr\'e, S\~ao Paulo, Brasil}

\begin{abstract}

An analytical solution of Einstein-Maxwell equations with a static fluid 
as a source is presented. The spacetime is represented by the axially symmetric 
Weyl metric and the  energy-momentum tensor describes a
coupling of a fluid  with an electromagnetic field.
 When appropriate limits are performed we 
recover the well-known solutions of Gutsunaev-Manko and Schwarzschild. 
Also, using Eckart's thermodynamics, we calculated the temperature, the 
mechanical pressure, the charge density and the energy density of the 
system. The analysis of  thermodynamic quantities suggests that the 
solution can be used to represent a magnetized compact stellar object 
surrounded by a charged  fluid.
\end{abstract}
\maketitle

\section{Introduction}

The study of magnetic fields in astrophysical objects such as white 
dwarfs, neutron stars, pulsars and black holes, has grown sharply in 
recent years \cite{akm:pan,fuk:war,sot:col,agu:pon,kun}. In fact, 
several observations show that there are various scenarios where the 
magnetic fields and general relativity can not be neglected . One of 
them is the presence of strong magnetic fields in active galactic nuclei 
\cite{lob,aha,zak:etal,gre:etal}. These nuclei are known to produce more 
radiation than the rest of the entire galaxy and directly affect its 
structure and evolution. Another scenario is the production of 
relativistic collimated jets in the inner regions of accretion discs, 
which can be explained considering magneto-centrifugal mechanisms 
\cite{liu:sha,aki:whe,mku,jet,ust:kol,kra:li}. Also, magnetic fields are 
important in understanding the interplay between magnetic and thermal 
processes for strongly magnetic neutron stars 
\cite{agu:pon,hen:fro,med:lai}. At least $10 \%$ of all neutron stars 
are born as magnetars, with magnetic fields above $10^{14}$~G 
\cite{li:van,pav:bez,ibr:swa}. Analytical models that describe these 
astrophysical objects are often associated with solutions of Einstein's 
equations \cite{sch1,sch2,ker,newman,Rei,nor}. In the context of 
relativistic hydrostatic models, the Tolman-Oppenheimer-Volkov 
\cite{tolman,oppi} equations describe the internal structure of general 
relativistic static perfect fluid spheres, e.g. neutron stars. In the 
search for more realistic models for compact stellar systems, the 
energy-momentum tensor, the source of  Einstein's equations, is modified by 
introducing more complex terms that take into account additional 
physical properties as, for example, electromagnetic fields. In the case 
of relativistic magneto-hydrodynamics the reduction of this 
 non-linear system  leads often to simple models with limited 
applicability. For this reason, they are frequently superseded by 
numerical models \cite{BHsim,GUZ,font}. Despite of this complexity, sometimes we 
can have useful analytical solutions, e.g., the Gutsunaev-Manko solution that
 describes the gravitational field of one static massive magnetic 
dipole \cite{gus}. The aim of this work is to construct an analytical 
solution to the Einstein-Maxwell field equations coupled with a fluid in order 
to represent a static configuration that can be used to characterize the 
gravitational field of  a  magnetized astrophysical 
object surrounded by a charged fluid.

The article is organized as follows. In Sec. \ref{two} we present the 
Einstein equations and the energy-momentum tensor to be considered. In Sec. 
\ref{three}, we exhibit  the solution of the Einstein equations. In Sec. 
\ref{four} we study the  thermodynamical properties of the system. 
Finally, in Sec. \ref{five}, we summarize our results.

\section{Coupling of  fields} \label{two}

The spacetime for our model is represented by the Weyl metric,
\begin{equation}
ds^{2}=f^{-1}\left( e^{2\Lambda }\left[ d\rho ^{2}+dz^{2}\right] +\rho
^{2}d\varphi ^{2}\right) -fdt^{2}, \label{eq:ds2}
\end{equation}

\noindent where $f = f(\rho, z)$ and $\Lambda = \Lambda(\rho, z)$.  The coordinate
 range are the usual for axial symmetry. Our 
conventions are: $G = c = 1$, metric signature +2, partial and covariant 
derivatives with respect to the coordinate $x_\mu$ denoted by $,\mu$ and 
$;\mu$, respectively. Greek indices run from 1 to 4, with $(1,2,3,4) = 
(\rho,z,\varphi,t)$, and Latin indices run from 1 to 3. The aim of this work is
 to solve Einstein's equations with an energy-momentum tensor 
representing the coupling between a fluid  and an electromagnetic 
field. Thus, in our model, the energy-momentum tensor is the sum of the 
electromagnetic energy-momentum tensor and a fluid energy-momentum 
tensor. The electromagnetic energy-momentum tensor considered is 
\begin{equation}
\overset{(EM)}{T_{\alpha\beta}} = \frac{1}{4\pi} \left[
F_{\alpha\mu}
F_{\beta}^{\;\;\mu} - \frac{1}{4} g_{\alpha \beta} F_{\mu\nu} F^{\mu\nu} 
\right], \label{eq:TEL}
\end {equation}
 
\noindent where $F_{\alpha\beta}$ is the electromagnetic field tensor 
defined as $F_{\alpha\beta}=A_{\alpha,\beta} - A_{\beta,\alpha}$, and 
$A_\mu$ is the four-potential $A_{\mu}= (U,V,W,\Phi)$, where the functions
$U,V,W,\Phi$ depend only on the coordinates 
$(\rho, z)$. The energy-momentum tensor of the fluid  is
\begin{equation}
\overset{(F)}{T_{\alpha\beta}} = \varepsilon v_{\alpha }v_{\beta}
+ (p-\varsigma
\theta )h_{\alpha \beta }+\tau_{\alpha \beta }, \label{eq:TNPF}
\end{equation}

\noindent where $v_{\alpha }$ represents the 4-velocity of the fluid, 
$\varepsilon$ is the fluid energy density, $p$ is the fluid pressure, 
$\varsigma $ is the bulk effective viscosity, $\theta = 
v^{\alpha}_{;\alpha}$ is the expansion, $\tau_{\alpha \beta}$ is the 
stress tensor defined as $\tau_{\alpha \beta }=-2\eta \sigma_{\alpha 
\beta} + q_{\alpha}v_{ \beta }+q_{\beta }v_{\alpha }$, $q^{\alpha }$ is 
the heat flux, $\eta$ is the shear viscosity, $h_{\alpha 
\beta}=v_{\alpha} v_{\beta}+g_{\alpha \beta }$ is the spatial projection 
tensor and $\sigma_{\alpha \beta}$ is the symmetric trace-free spatial 
shear tensor given by
\begin{equation}
\sigma_{\alpha \beta }=\frac{1}{2}\left\{(v_{\alpha }h_{
\text{ \ }\beta }^{\mu })_{;\mu}+ (v_{\beta }h_{\text{ \
}\alpha }^{\mu })_{;\mu}\right\} -\frac{1}{3}\theta h_{\alpha \beta }.
\label{eq:Sigma}
\end{equation}

The Einstein's equations for the system fluid plus
 electromagnetic field are
\begin{equation}
G_{\alpha \beta }=8\pi \left[ \overset{(EM)}{T_{\alpha\beta}}
+ \overset{(F)}{T_{\alpha\beta}} \right].\label{eq:einstein}
\end{equation}

\noindent In the next section we solve Eq. (\ref{eq:einstein}), with 
$\overset{(EM)}{T_{\alpha\beta}}$ and $\overset{(F)}{T_{\alpha\beta}}$ 
given by (\ref{eq:TEL}) and (\ref{eq:TNPF}) respectively.

\section{A solution to the  Einstein's equations} \label{three}

To solve the Einstein equations we use  a  reference frame
co-moving  with  the fluid. In this reference frame, 
the four-velocity of the fluid is $v_{\alpha} =\left[0,0,0,v_{4} 
=\sqrt{f} \right]$. Note also that $\theta = v^{\alpha}_{;\alpha}=0$ in 
this frame, i.e. the expansion of the fluid is null and there is no 
divergence or convergence of the fluid world lines. For this reason a 
co-moving observer does not see an effective spatial electric current. Therefore, 
in our co-moving frame we have that the electric current is null,  $J^i=0$. The 
condition $J^1 = J^2 = 0$, together with Maxwell equations,
\begin{equation}
4 \pi J^{\mu }=\nabla _{\alpha }F^{ \mu \alpha }, \label{eq:JOTA}
\end{equation}

\noindent leads to
\begin{equation}
\rho fe^{-2\Lambda }\left( V_{,\rho }-U_{,z}\right)=\kappa_{0},
\label{eq:c1}
\end{equation}

\noindent where $\kappa_{0}$ is a constant, whereas the condition 
$J^{3}= 0$ can be expressed by
\begin{equation}
\overrightarrow{\nabla }\cdot \left[ \frac{f}{\rho 
}\overrightarrow{\nabla }W\right] =0, \label{eq:JOTA3}
\end{equation}

\noindent where $\overrightarrow{\nabla} \equiv \hat{\rho} 
\frac{\partial} {\partial \rho}+\hat{z} \frac{\partial}{\partial z}$. 
Also, with the help of $J^4$ obtained from (\ref{eq:JOTA}), we can write 
the charge density of the system $d_c=-J_\mu v^\mu$ as
\begin{eqnarray}
d_c = \frac{f^3}{4 \pi \rho e^{2\Lambda}} \overrightarrow{\nabla} \cdot 
\left[ \frac{\rho}{f} \overrightarrow{\nabla} \Phi \right]. \label{eq:dc}
\end{eqnarray}

\noindent Another simple equation 
can be obtained from the components \{1,3\} and \{2,3\} of 
Einstein's equations. Both components are equal to zero and can be 
written as
\begin{equation}
(V_{,\rho} - U_{,z}) W_{,z} = (V_{,\rho} - U_{,z}) 
W_{,\rho} = 0. \label{eq:comp}
\end{equation}

\noindent One  possible way to satisfy Eq. (\ref{eq:comp}) is to 
set the magnetic potential $W = {\rm constant}$. With this condition 
Eqs. (\ref{eq:JOTA3}) and (\ref{eq:comp}) are satisfied, but this leads 
to a physical model with tensions instead of pressures, and moreover 
these tensions are independent of the coordinate $z$. So, we discarded 
this solution. Another possibility is to set the magnetic potential $W 
\neq {\rm constant}$. With this last condition Eqs. (\ref{eq:comp}) are 
satisfied when
\begin{equation}
V_{,\rho} - U_{,z} = 0. \label{VU}
\end{equation}

\noindent Therefore, the constant $\kappa_0$ in (\ref{eq:c1}) is equal 
to zero. Now, if we add Einstein's equations \{1,1\} and \{2,2\} we find 
that the fluid pressure is equal to
\begin{equation}
p=-\frac{f^{2}}{4 \pi e^{4\Lambda}} \left( V_{,\rho} - U_{,z} \right),
\label{pres}
\end{equation}

\noindent from which we obtain, using Eq. (\ref{VU}), that it is equal 
to zero. From these considerations, the Einstein equations reduce to two 
parts, one integrable system for the functions $f$, $W$, $\Phi$ and 
$\Lambda$ given by
\begin{eqnarray}
&&\Lambda _{,\rho }=\frac{\rho}{4}\frac{\left( f_{,\rho
}^{2}-f_{,z}^{2}\right)}{f^{2}}+\frac{f}{\rho}\left( W_{,\rho}^2 
-W_{,z}^2 \right) \nonumber \\
&&\hspace{1cm}-\frac{\rho}{f}\left( 
\Phi_{,\rho}^2-\Phi_{,z}^2\right), \label{eq:eq1} \\
&&\Lambda _{,z}=\frac{\rho f_{,z}f_{,\rho }}{2f^{2}}+\frac{2f}{\rho
}W_{,\rho }\ W_{,z}-\frac{2 \rho}{f}\Phi_{,\rho }\ \Phi_{,z},
\label{eq:eq2}\\
&&\Lambda _{,z,z}+\Lambda _{,\rho ,\rho }=-\frac{1}{4}\frac{f_{,\rho
}^{2}+f_{,z}^{2}}{f^{2}} +\frac{f}{\rho ^{2}}\left( W_{,\rho}^2
+W_{,z}^2 \right) \nonumber \\
&&\hspace{2.3cm}+\frac{1}{f}\left( \Phi_{,\rho}^2
+\Phi_{,z}^2\right), \label{eq:eq3} \\
&&0=\overrightarrow{\nabla }\cdot
\left[ \frac{f}{\rho }\overrightarrow{\nabla } W\right], \label{eq:eq4}
\end{eqnarray}

\noindent and another system of equations for the fluid density energy 
and flux radiation, which can be written as
\begin{eqnarray}
\varepsilon &=& \frac{f}{4 \pi \rho e^{2\Lambda}} \beta 
,\label{eq:qq44}\\
q_{1}&=&0 ,\label{eq:qq11}\\
q_{2}&=&0 ,\label{eq:qq22}\\
q_{3}&=&-\frac{\sqrt{f}}{4\pi e^{2\Lambda}}\left[
\overrightarrow{\nabla }W \cdot \overrightarrow{\nabla }
\Phi \right],\label{eq:qq33}
\end{eqnarray}

\noindent where $\beta$ is defined by
\begin{equation}
\beta = \frac{1}{2}\overrightarrow{\nabla }\cdot \left( \frac{\rho }{f
}\overrightarrow{\nabla }f\right) -\frac{f}{\rho } 
\overrightarrow{\nabla }
W\cdot \overrightarrow{\nabla }W-\frac{\rho }{f}\overrightarrow{\nabla }
\Phi\cdot \overrightarrow{\nabla }\Phi .\label{eq:beta}
\end{equation}

\noindent In the next subsection we find a solution for $f$, $W$, $\Phi$ 
and $\Lambda$.

\subsection{Solution to the integrable system}

In order to solve the system of equations (\ref{eq:eq1}-\ref{eq:eq4}), 
we first use Eqs. (\ref{eq:eq1}) and (\ref{eq:eq2}) to find the 
integrability condition $\Lambda_{,\rho,z}=\Lambda_{,z,\rho}$. This 
condition can be written as
\begin{equation}
\frac{f_{,z}}{f} \beta=2\Phi_{,z} \overrightarrow{\nabla }\cdot \left(
\frac{\rho}{f} \overrightarrow{\nabla}\Phi\right).\label{eq:ef1}
\end{equation}
\noindent Furthermore, if we substitute Eqs. (\ref{eq:eq1}) and 
(\ref{eq:eq2}) into (\ref{eq:eq3}) we obtain that
\begin{equation}
\frac{f_{,\rho }}{f}\beta=2\Phi_{,\rho } \overrightarrow{\nabla }\cdot
\left(\frac{\rho}{f} \overrightarrow{\nabla}\Phi\right).\label{eq:ef2}
\end{equation}
\noindent Assuming that the fluid energy density ($\varepsilon$) and 
charge density of the system ($d_c$) are different from zero, which 
implies from Eqs. (\ref{eq:dc}) and (\ref{eq:qq44}) that $\beta$ and 
$\overrightarrow{\nabla }\cdot \left(\frac{\rho}{f} 
\overrightarrow{\nabla}\Phi\right)$ are also different from zero, we 
find from Eqs. (\ref{eq:ef1}) and (\ref{eq:ef2}) that
\begin{equation}
f_{,\rho }\Phi_{,z}=f_{,z}\Phi_{,\rho } \mbox{ ,}
\end{equation}
\noindent which is satisfied when $\Phi=\Phi\left( 
f\right)$. Using this last condition, we can rewrite equation 
(\ref{eq:ef1}) as
\begin{eqnarray}
\left( 2f\Phi_{,f}-
\Phi\right) \overrightarrow{\nabla }\cdot \left( \frac{\rho }{f}
\overrightarrow{\nabla }\Phi\right)= \frac{1}{2}\overrightarrow{\nabla
}\cdot \left( \frac{\rho }{f} \overrightarrow{\nabla }f\right) \nonumber 
\\
-\frac{f}{\rho }\overrightarrow{\nabla }
W\cdot \overrightarrow{\nabla }W-\overrightarrow{\nabla }\cdot  \left(
\frac{
\rho }{f}\Phi\overrightarrow{\nabla }\Phi\right)
.\label{eq:fundamental}
\end{eqnarray}

 The explicit functional form of $\Phi(f)$ is 
obtained in Appendix A, replacing the function $f$ by a power series 
of $\Phi$ in order to satisfy Eq. (\ref{eq:eq4}). The result is 
$\Phi=\Phi_{0} \sqrt{f}$, where $\Phi_0$ is a constant. Now, 
substituting the functional form of $\Phi$ into Eq. 
(\ref{eq:fundamental}) we obtain
\begin{eqnarray}
\frac{1-\Phi_0^2}{2} \overrightarrow{\nabla} \cdot \left( \frac{\rho}{f} 
\overrightarrow{\nabla} f \right) - \frac{f}{\rho} 
\overrightarrow{\nabla} W \cdot
\overrightarrow{\nabla} W = 0 , \label{eq:phi0}
\end{eqnarray}
\noindent which has the same form of the Gutsunaev-Manko's equation for 
the massive magnetic dipole in  vacuum if we define 
$W=\sqrt{1-\Phi^2_0} \mathcal{A}$, where $\mathcal{A}$ is the 
electromagnetic potential considered by Gutsunaev and Manko \cite{gus}. 
The solution of Eqs. (\ref{eq:eq4}) and (\ref{eq:phi0}) written in 
prolate ellipsoidal coordinates $x=(r_{+} + r_{-})/2k$ and $y=(r_{+} - 
r_{-})/2k$, where $r_{\pm} = \sqrt{\rho^2+(z\pm k)^2}$ and $k$ is a real 
constant, is studied  in the Appendix B. With this solution, the functions 
$f$, $\Phi$, $W$ and $\Lambda$ can be written in terms of only one 
parameter $\alpha$ as
\begin{widetext}
\begin{eqnarray}
f&=&\frac{x-1}{x+1}\left(\frac{\left[x^{2}-y^{2}+\alpha^{2}(x^{2}-1)
\right]^{2}
+4 \alpha^{2}
x^{2}(1-y^{2})}{\left[x^{2}-y^{2}+\alpha^{2}(x-1)^{2}\right]^{2}
-4 \alpha^{2}y^{2}(x^{2}-1)} \right)^{2} ,\label{eq:solf}\\
\Phi&=&\Phi_{0}\sqrt{\frac{x-1}{x+1}}\left(\frac{\left[x^{2}-y^{2}+
\alpha^{2}(x^{2}-1)\right]^{2}
+4 \alpha^{2}
x^{2}(1-y^{2})}{\left[x^{2}-y^{2}+\alpha^{2}(x-1)^{2}\right]^{2}
-4 \alpha^{2}y^{2}(x^{2}-1)} \right) ,\label{eq:solfi}\\
W&=&\frac{4k
\alpha^{3}\sqrt{1-\Phi_{0}^{2}}\left(1-y^{2}\right)\left[2\left(
\alpha^{2}+1\right)x^{3}+\left(1-3\alpha^{2}\right)x^{2}+y^{2}+
\alpha^{2}\right]}
{\left(\alpha^{2}+1\right) \left( \left[x^{2}-y^{2}+\alpha^{2}
\left(x^{2}-1 \right) \right]^{2}+4\alpha^{2} x^{2} \left(1-y^{2}\right) \right)} ,\label{eq:solW}\\
e^{2\Lambda}&=&\left[\frac{\left(x^{2}-1\right)}{(\alpha^{2}+1)^{8}}
\frac{\left(
\left[x^{2}-y^{2}+\alpha^{2}(x^{2}-1)\right]^{2}+4
\alpha^{2}x^{2}(1-y^{2})\right)^{4}}
{(x^{2}-y^{2})^9}\right]^{(1-\Phi_{0}^{2})} \mbox{.} \label{eq:solL}
\end{eqnarray}
\end{widetext}
\noindent In the next subsection we relate the parameter $\alpha$ with 
the magnetic dipole of the system.

\subsection{Asymptotic solution}

Let us study the behavior of the metric and the electromagnetic fields 
far from the source. In this case, it is appropriate to write the 
functions $f$, $\Phi$, $W$ and $\Lambda$ in spherical coordinates and 
 expand them in power series in $r^{-1}$. The spherical coordinates 
are related to the prolate spherical coordinates through the expressions 
$x = (r-m)/k$ and $y=\cos \theta$, where $m$ is a real parameter. 
The 
function $f$ far from the source takes the form
\begin{eqnarray}
&&\hspace{-0.5cm} f = 1+ \frac{2k (-1+3 \alpha^2)}{(1+\alpha^2) r} 
\nonumber \\
&&+ \frac{2 k 
(-1+3\alpha^2) (-k + m + 3k\alpha^2 + m \alpha^2)}{(1+\alpha^2)^2 r^2} + 
O(r^-3). \nonumber \\  \label{eq:fserie}
\end{eqnarray}
\noindent Now, we imposed to the function $f$ to have a Schwarzschild form 
far from the source, we find
\begin{eqnarray}
k=\frac{m(1+\alpha^2)}{1-3\alpha^2}.
\end{eqnarray}
\noindent Note that with this form of $k$, the third term in Eq. 
(\ref{eq:fserie}) is equal to zero, so finally we obtain that $f=1-2m/r 
+ O(r^{-3})$. The function $e^{2\Lambda}$ goes as $e^{2\Lambda}=1 + 
O(r^{-2})$. Therefore, far from the source we obtain the
Schwarzschild metric. Hence  $m$ represents the mass of our system.

Now, we analyze the asymptotic behavior of the electromagnetic 
potential, $A_\mu = [0,0,W,\Phi]$. For this purpose, we use the 
components of the four-potential ($W,\Phi$) written in spherical 
coordinates. Expanding Eqs. (\ref{eq:solfi}) and (\ref{eq:solW}) in 
series of $r^{-1}$ we obtain that
\begin{eqnarray}
&&W = \frac{8 \sqrt{1-\Phi^2_0} m^2 \alpha^3 \sin^2
\theta}{(1-3\alpha^2)^2 r} + O(r^{-2}), \label{eq:Wserie} \\
&&\Phi = \Phi_0 - \frac{\Phi_0 m}{r} + O(r^{-2}). \label{eq:fiserie}
\end{eqnarray}
\noindent Comparing Eq. (\ref{eq:Wserie}) with the classical magnetic 
potential, we note that the magnetic dipole moment is equal to
\begin{eqnarray}
\mu = \frac{8 \sqrt{1-\Phi^2_0} m^2 \alpha^3}{(1-3\alpha^2)^2},
\label{eq:momento}
\end{eqnarray}
\noindent which differs from the Gutsunaev-Manko magnetic dipole by a 
factor of $\sqrt{1-\Phi_0^2}$. From Eq. (\ref{eq:momento}) we can relate 
the parameter $\alpha$ with the magnetic dipole moment. Note that the 
parameter $\Phi_0$ can take values between -1 and 1. Therefore, from 
Eqs. (\ref{eq:Wserie}) and (\ref{eq:fiserie}), we see that for values of 
$|\Phi_0|$ near zero, the magnetic potential $W$ and the electric 
potential $\Phi$ far from the source attained their higher and lower 
values respectively. For values of $|\Phi_0|$ near one, we have the 
opposite behavior. Note that the fluid total charge is $m\Phi_0$.

\section{Thermodynamic properties } \label{four}

%%%%%%%%%%%%%%%%%%%%%%%%%%%%%%%%%%%%%%%

\begin{figure*} \vspace{0.7cm}
\epsfig{width=6.5cm, height=6.cm, file=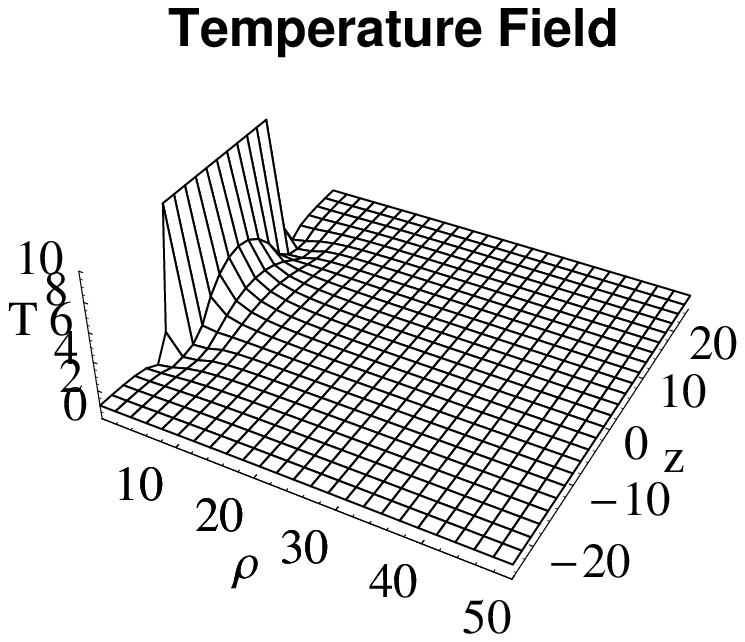} \hspace{1cm} 
\epsfig{width=6.5cm, height=6.cm, file=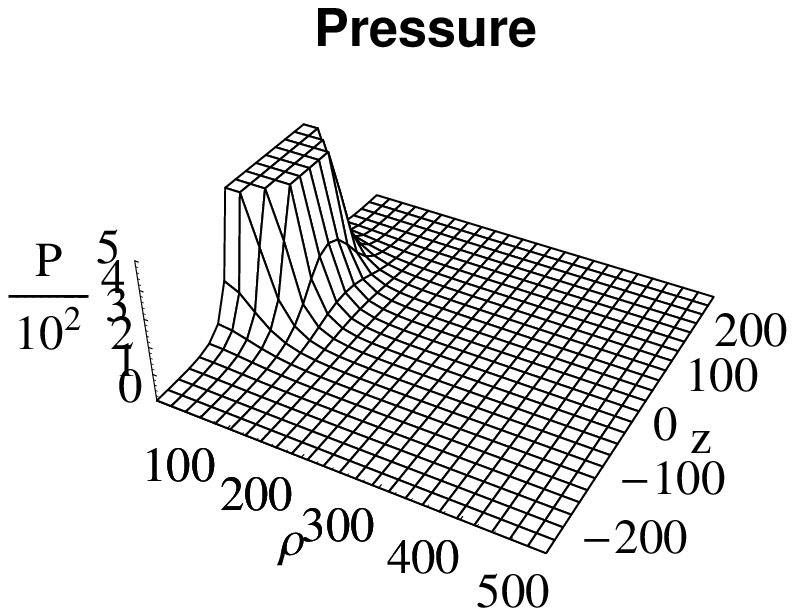}

\vspace{0.5cm} 

\epsfig{width=6.5cm, height=6.cm, file=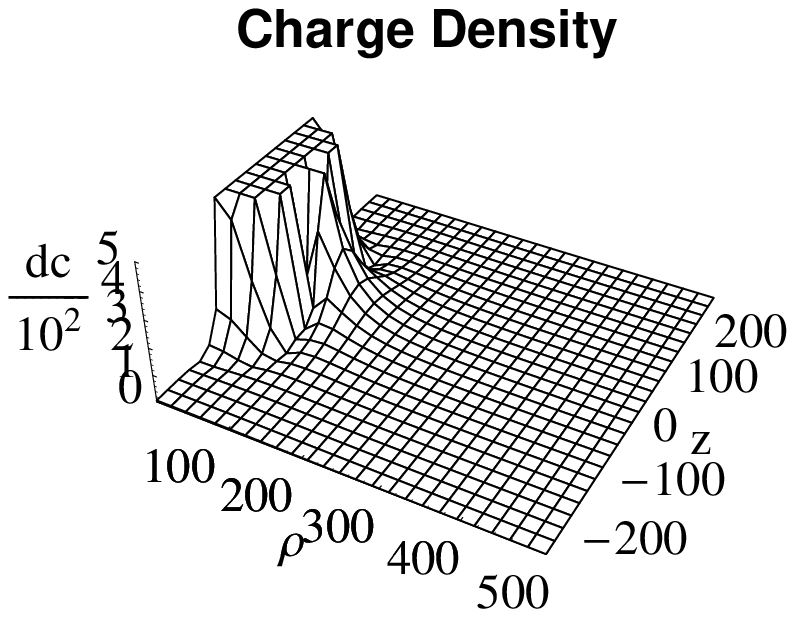} \hspace{1cm} 
\epsfig{width=6.5cm, height=6.cm, file=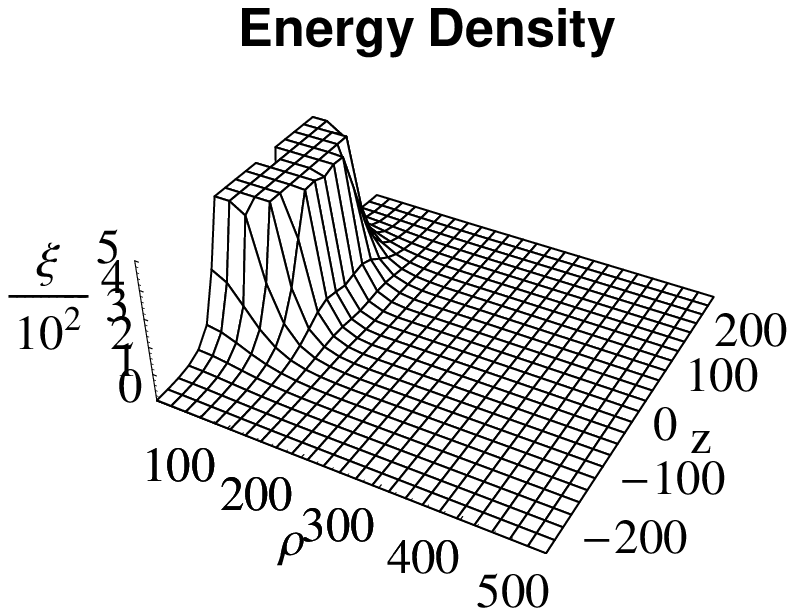}

\caption{Profiles of the temperature, the mechanical pressure, the 
charge density and the energy density of the system. The values of the 
parameters are $T_\infty=1$, $m=1$, $\alpha=0.54$ and $\Phi_0=0.5$.} 
\label{fig1}

\end{figure*}

%%%%%%%%%%%%%%%%%%%%%%%%%%%%%%%%%%%%%%%

A thermodynamic analysis of the system can be made performing a 
decomposition of the energy-momentum tensor $T_{\alpha\beta} = 
\overset{(EM)}{T_{\alpha\beta}} + \overset{(F)}{T_{\alpha\beta}}$ in its 
proper components as made by Eckart \cite{Eck}. Note that for static system 
Eckart's  thermodynamics does not present causality problems. Following Eckart, 
we can write the energy-momentum tensor in its proper 
components as
\begin{eqnarray}
T^{\alpha \beta} = \xi v^\alpha v^\beta + Q^\alpha v^\beta + Q^\beta
v^\alpha + \lambda^{\alpha\beta},
\end{eqnarray}
\noindent with $\xi$, $Q^\alpha$, $\lambda^{\alpha \beta}$ are given by
\begin{eqnarray}
&&\xi= v_\alpha v_\beta T^{\alpha\beta}, \label{eq:xi}\\
&&Q^\alpha = - h^\alpha_\beta T^{\beta\gamma} v_{\gamma}, \label{eq:qs}\\
&&\lambda^{\alpha\beta} = h^\alpha_\gamma h^\beta_\delta
T^{\gamma\delta}.\label{eq:strees}
\end{eqnarray}
\noindent where $\xi$ is the energy density of the system, $Q^\alpha$ is 
the energy flux of the system and $\lambda^{\alpha\beta}$ is the stress 
tensor of the system. Calculating explicitly the components of the 
energy flux of the system, with the help of Eqs. (\ref{eq:qq11}) and 
(\ref{eq:qq33}), we found that $Q^1=Q^2=Q^3=0$ which means that the 
system is in an equilibrium configuration. In this case the temperature 
of the system obeys the relation \cite{Eck,is2,his},
\begin{eqnarray}
h^{\alpha\beta} [ \nabla_\beta T + T v^\mu \nabla_\mu v_\beta ] =0.
\end{eqnarray}
\noindent For $\alpha=1,2,3$ we have that
\begin{eqnarray}
&&\partial_\rho \ln \left[ T \sqrt{f} \right] = 0, \label{eq:t11}\\
&&\partial_z \ln \left[ T \sqrt{f} \right] = 0, \label{eq:t22} \\
&&\partial_\varphi T = 0. \label{eq:t33}
\end{eqnarray}
\noindent Using the above equations we state that the temperature of the 
system is equal to
\begin{eqnarray}
T(\rho,z) = \frac{T_\infty}{\sqrt{f}},\label{eq:temp}
\end{eqnarray}
\noindent where $T_\infty$ is the temperature at infinity.

The mechanical pressure of the system ($P$) is given in Eckart's 
thermodynamics by the expression
\begin{eqnarray}
P &=& \frac{1}{3} g_{\alpha\beta} \lambda^{\alpha\beta} = \frac{1}{3}
g_{\alpha\beta} h^\alpha_\gamma h^\beta_\delta T^{\gamma\delta}
\nonumber \\
&=& \frac{(1-\Phi_0^2)f^N}{48 \pi \rho e^{2\Lambda}}
\overrightarrow{\nabla } \cdot \left( \frac{\rho}{f^N}
\overrightarrow{\nabla } f \right),
\end{eqnarray}
\noindent where $N=(2-3 \Phi_0^2)/(2-2\Phi_0^2)$. The energy density of 
the system (\ref{eq:xi}) can be written as
\begin{equation}
\xi=\frac{\left(1+\Phi_{0}^{2}\right) f^{M}}{16 \pi \rho
e^{2\Lambda}}  \overrightarrow{\nabla }\cdot \left( \frac{\rho
}{f^{M} }\overrightarrow{\nabla }f\right),
\end{equation}
\noindent where $M = (2+3\Phi_0^2)/(2 + 2 \Phi_0^2)$.

In Fig. \ref{fig1} we present a graph of the temperature, the mechanical 
pressure, the charge density and the energy density of the system. Note 
that all variables decay rapidly to their asymptotic values, which 
suggest that we can treat our system as a compact stellar object. While 
varying the values of the parameters $\Phi_0$, $\alpha$ and $m$ we found 
that we can obtain models which are less or more compact. For higher 
values of $\Phi_0$ we obtain the more compact objects.

\section{Conclusions} \label{five}

In this work we found an analytical axially symmetric static solution of 
Einstein's equations with a energy-momentum tensor which couples an 
electromagnetic field and a fluid field. Far from the source, our metric 
is consistent with the Schwarzschild solution. Moreover, if we let 
$\Phi_0 = 0$ we recover the Gutsunaev-Manko solution. The thermodynamic 
variables studied in our model suggest that we can treat our model as a 
compact stellar object, and these variables also may allow a direct 
comparison with  observations. 

The solution to the Einstein equations presented in this work is a  static 
metric. This looks like  to be in contradiction to previous results that 
state that in the spacetime associated  to a  static magnetic dipole in the 
presence of electric charges appears a frame-dragging effect 
\cite{bon,man:rod}, i.e.  we have a stationary metric. In our case, the static 
configuration is possible because the heat flux and the electromagnetic energy
flux (\ref{eq:qq33}) compensate each other in such a way that the total 
energy flux is equal to zero. Static metrics associated to the 
equilibrium configurations obtained from the 
superposition of  counterrotating fluxes in the context of General Relativity 
are not new, e.g, in the Morgan \& Morgan disks \cite{mor:mor} we have the
 same number of particles rotating in opposite directions (counterrotating
hypothesis).

We conclude by  mentioning that our solution that represents a charged fluid 
around a massive magnetic dipole 
 may be useful to model a magnetar within a fluid.

\section{Acknowledgment}

J.D.P. thanks CNPQ for financial support and C. Dobrigkeit for valuable 
suggestions; P.S.L.  thank FAPESP and CNPq for financial support.

\section*{APPENDIX A: FUNCTIONAL FORM OF $\Phi(f)$} \label{appendixA}

To find the functional form of $\Phi$ first we write the function $f$ as 
a power series of $\Phi$
\begin{equation} 
f = \sum_{n=0}^{\infty} a_n \Phi^n, \label{serie} 
\end{equation}
\noindent where $a_n$ are real coefficients. Replacing (\ref{serie}) in 
(\ref{eq:eq4}), we obtain
\begin{eqnarray}
\sum_{n=0}^{\infty} a_n \overrightarrow{\nabla} \cdot \left[ 
\frac{\Phi^n}{\rho} 
\overrightarrow{\nabla} W \right] = a_0 \overrightarrow{\nabla} \cdot 
\left[ 
\frac{1}{\rho} 
\overrightarrow{\nabla} W \right] \nonumber \\
+ a_1 \overrightarrow{\nabla} \cdot \left[ \frac{\Phi}{\rho}
\overrightarrow{\nabla} W \right] + \dots = 0. \label{eq:serie2}
\end{eqnarray}

To simplify the notation we define the quantities $B_n$ as
\begin{equation}
B_n = \overrightarrow{\nabla} \cdot \left[ \frac{\Phi^n}{\rho} 
\overrightarrow{\nabla} W 
\right],
\end{equation}
\noindent with $n=0,1,2,3,\dots, \infty$, which satisfy the recurrence 
relation
\begin{equation}
B_n = 2 \Phi B_{n-1} - \Phi^2 B_{n-2}. \label{eq:recurrence}
\end{equation}

With the help of Eq. (\ref{eq:recurrence}) we can cast Eq. 
(\ref{eq:serie2}) into the form
\begin{eqnarray}
\sum_{n=0}^\infty a_n B_n = B_1 \sum_{n=1}^\infty (a_n - 2 a_{n+2}) n 
\Phi^n \nonumber \\ 
+ B_2 \sum_{n=2}^\infty (n-2) a_n \Phi^{n-2}=0. \label{eq:serie3}
\end{eqnarray}

To obtain one condition from Eq. (\ref{eq:serie3}) to help us to find 
the functional form $\Phi(f)$, we demand that $a_n = 2 a_{n+2}$ for all 
$n \geq 1$. This allows us to write the above series only in terms of 
$B_2$, say
\begin{eqnarray}
\sum_{n=0}^\infty a_n B_n = \frac{1}{2} B_2 \Phi^2 \left[ a_1 
\sum_{n=0}^\infty \frac{2n+1}{2^n} \Phi^{2n} \nonumber \right. \\
\left. + a_2 \sum_{n=0}^\infty \frac{n+1}{2^n} \Phi^{2n+1} \right] = 0.
\end{eqnarray}
\noindent The above equation is satisfied if we set $B_2=0$, so
\begin{eqnarray}
B_2 = \overrightarrow{\nabla} \cdot \left[ \frac{\Phi^2}{\rho}
\overrightarrow{\nabla} W \right] = 0. \label{eq:b2}
\end{eqnarray}
\noindent Finally, by direct comparison between Eq. (\ref{eq:b2}) and 
Eq. (\ref{eq:eq4}) we obtain that the functional form of $\Phi$ is $\Phi 
= \Phi_0 \sqrt{f}$, where $\Phi_0$ is a constant. Note that by 
D'Alembert criterion the series of Eq. (\ref{serie}) converges when 
$|\Phi| < \sqrt{2}$.

\section*{APPENDIX B: SOLUTION OF EQS. (\ref{eq:JOTA3}) AND 
(\ref{eq:phi0})} \label{appendixB}

Using the prolate ellipsoidal coordinates defined as
\begin{equation}
\rho=k \sqrt{x^{2}-1}\sqrt{1-y^{2}}, \;\;\;  z=kxy ,
\end{equation}
\noindent where $k$ is a real parameter, we can write Eqs. 
(\ref{eq:eq4}) and (\ref{eq:phi0}) in the form
\begin{eqnarray}
&&\left( \frac{f}{1-y^2} W_{,x} \right)_{,x} + \left( 
\frac{f}{x^2 -1} W_{,y} \right)_{,y} = 0, \label{eq:devagar1} \\
&&\left(\frac{x^2-1}{f} f_{,x} - \frac{2\left(1-\Phi^2_0 \right)}{k^2}
\frac{Wf}{1-y^2} W_{,x} \right)_{,x} \nonumber \\
&& + \left(\frac{1-y^2}{f} f_{,y} 
- \frac{2 \left(1-\Phi^2_0 \right)
}{k^2} \frac{Wf}{x^2-1} W_{,y} \right)_{,y} = 0. \label{eq:devagar2}
\end{eqnarray}
\noindent Now, we define the auxiliary functions $H$ and $G$, so that 
$H=\frac{\sqrt{1-\Phi_0^2}}{k (1-y^2)} W$ and $G = \frac{x+1}{x-1} f$. 
With these definitions, Eq. (\ref{eq:devagar1}) is trivially satisfied 
by a function $I(x,y)$ if
\begin{eqnarray}
&&I_{,y} = \frac{x-1}{x+1} G H_{,x}, \label{eq:omega1} \\
&&I_{,x} = - \frac{G}{(x+1)^2} [ (1-y^2) H]_{,y}. \label{eq:omega2}
\end{eqnarray}
\noindent So, Eq. (\ref{eq:devagar2}) can be written in terms of $G$, 
$H$ and $I$, say
\begin{eqnarray}
\lefteqn{\left(I_{,yy}-I_{,xx}\right)H=} \nonumber \\
& &\frac{ x^{2}-y^{2}}{1-y^{2}} \left( \left[ \frac{G_{,x}}{2G}\right]
_{,y}+ \frac{\left( x+1\right)I_{,x} I_{,y} }{G\left(
x-1\right)}\right). \label{eq:fi3}
\end{eqnarray}
\noindent The next step is to assume that $G$, $H$ and $I$ are functions 
of the form
\begin{equation}
I =\frac{N}{D}, \mbox{\hspace{0.5cm}} H=\frac{S}{M}
\mbox{\hspace{0.5cm}and \hspace{0.5cm}} G=\frac{M^2}{D^2},\label{eq:fra}
\end{equation}
\noindent where $N,D,S$ and $M$ are  polynomials in $x$ and $y$ of the form,
\begin{eqnarray}
N(x,y) = \sum_{j=0}^{j_{max}} \sum_{m=0}^{j_{max}-j} \eta_{jm} x^j y^m \\
S(x,y) = \sum_{j=0}^{j_{max}} \sum_{m=0}^{j_{max}-j} \sigma_{jm} x^j y^m
\\
D(x,y) = \sum_{j=0}^{j_{max}} \sum_{m=0}^{j_{max}-j} \delta_{jm} x^j y^m
\\
M(x,y) = \sum_{j=0}^{j_{max}} \sum_{m=0}^{j_{max}-j} \mu_{jm} x^j y^m 
\end{eqnarray}
\noindent with $\eta_{jm}$, $\sigma_{jm}$, $\delta_{jm}$, $\mu_{jm}$ 
being  unknown coefficients to be determined. Substituting the 
relations (\ref{eq:fra}) into Eqs. (\ref{eq:omega1}) and 
(\ref{eq:omega2}) we obtain the set of equations
\begin{eqnarray}
&&(x+1) (D N_{,y} - N D_{,y} ) - (x-1) (M S_{,x} - S M_{,x} ) = 0,
\nonumber \\
&&(x+1)^2 (D N_{,x} - N D_{,x}) + M^2 \left[ (1-y^2) \frac{S}{M} 
\right]_{,y} =0. \nonumber \\ \label{eq:devagar3}
\end{eqnarray}

The system of Eqs (\ref{eq:devagar3}), together with Eq. (\ref{eq:fi3}) 
can be solved in a tedious direct comparison of the coefficients of the 
polynomials involved. The solution depends on the number of coefficients 
for each polynomial, i.e. the value of $j_{max}$. Furthermore, we  impose that
 the solution obtained be symmetric on the plane $z=0$ and also that has
the Schwarzschild solution as a particular case. The first physical 
solution is obtained when $j_{max}=4$. In this case the coefficients can 
be written in terms of only one parameter, $\alpha$. The solution is
\begin{eqnarray}
&&N=8\alpha^{3} xy \left(x-1\right) , \nonumber \\
&&D=\left[x^{2}-y^{2}+\alpha^{2}(x-1)^{2}\right]^{2}-4
\alpha^{2}y^{2}(x^{2}-1),  \nonumber \\
&&M=\left[x^{2}-y^{2}+\alpha^{2}(x^{2}-1)\right]^{2}
+4 \alpha^{2}
x^{2}(1-y^{2}) , \nonumber \\
&&S=4 \alpha^{3}
\left[2x^{3}+\frac{\left(1-3\alpha^{2}\right)}{\left(\alpha^{2}
+1\right)}x^{2}+\frac{y^{2}+\alpha^{2}}{\left(\alpha^{2}+1\right)}
\right] .\label{eq:SS}
\end{eqnarray}
\noindent Using the solution (\ref{eq:SS}), the functions $f$, $\Phi$ 
and $W$ are known. The function $\Lambda$ is calculated integrating Eqs. 
(\ref{eq:eq1}) and (\ref{eq:eq2}).

\end{document}